# Field-induced giant static dielectric constant in nano-particle aggregates at room temperature

F. CHEN[1], J. SHULMAN[1], S. TSUI[1], Y. Y. XUE[1], W. WEN[2], P. SHENG[2] and C. W. CHU[*1,3,4]

[1]*Department of Physics and Texas Center for Superconductivity, University of Houston, 202 Houston Science Center, Houston, Texas 77204-5002*

[2]*Department of Physics and Institute of Nano Science and Technology, Hong Kong University of Science and Technology, Clear Water Bay, Kowloon, Hong Kong*

[3]*Hong Kong University of Science and Technology, Clear Water Bay, Kowloon, Hong Kong*

[4]*Lawrence Berkeley National Laboratory, 1 Cyclotron Road, Berkeley, California 94720*

The analogy between magnetism and electricity has long been established by Maxwell in the 19$^{th}$ century, in spite of their subtle difference. While magnetic materials display paramagnetism, ferromagnetism, antiferromagnetism, and diamagnetism, only paraelectricity, ferroelectricity, and antiferrolelectricity have been found in dielectric materials. The missing 'diaelectricity' may be found if there exists a material that has a *dc*-polarization opposing the electric field or a negative dielectric susceptibility ε'-1, with ε' being the real part of the relative dielectric constant. Both of these properties have been observed in nano-particle aggregates under a dc electric bias field at room temperature. A possible collective effect in the nano-particle aggregates is proposed to account for the observation. 'Diaelectricity' implies overscreening by polarization to the external charges. Materials with a negative static ε' are expected to provide attraction to similar charges and unusual scattering to electromagnetic waves with possible profound implications for high temperature superconductivity and communication.

*Keywords: negative dielectric constant, negative polarization, diaelectricity, nano-particles, electrorheology*

A negative static dielectric constant, ε'(k,ω), at wave-vector k ≠ 0 and frequency ω = 0, has attracted great interest because of its implications for high temperature superconductivity, wave propagation, electrolyte behaviour, colloidal properties, and bio-membrane functions [1-4]. Its existence in metals and their associated plasmas has been predicted [1,2]. Unfortunately, negative ε'(k,ω=0) has not been reported in these systems. The extremely large ratio between the conductivity, σ(k,ω=0), and the dielectric admittance, ωε'(k,ω=0), makes the measurement practically impossible [3]. However, artificial structures have been recently proposed to modify the plasma parameters and lead to a negative ε'(k,ω) [3,4], albeit in the microwave region. One obstacle a system with a negative ε'(k,ω=0) must overcome is instability. It is thought that instability can lead to phase separation, and thus, a destruction of the state. However, a state of negative compressibility in 2D carrier gasses has recently been shown to survive such a phase separation under rather broad conditions [5].

A giant electric-field-induced yield-stress has also been reported [6,7] in an electrorheological (ER) fluid composed of a colloidal suspension of urea-coated nano-particles of $Ba_{0.8}Rb_{0.4}TiO(C_2O_4)_2$ (U-BRTOCO) in silicone oil, suggesting an unusual field effect on the ε' of nano-structure materials. We have therefore searched for such an effect on the ε'(ω) of the U-BRTOCO nano-particle aggregates.

The complex dielectric constant ε(ω)$_{k≠0}$ of the U-BRTOCO nano-particle ER fluid samples [6] was measured at room temperature under a dc bias field ($E_b$) between two plate-electrodes. ε(ω) ≡ ε'(ω) − iε''(ω) = ε'(ω) − i[σ(ω) − σ(0)]/ω = − i$I_{ac}$d/(ω$V_{ac}$S), in the Heaviside-Lorentz units, is determined by measuring the ac current $I_{ac}$ through and the ac voltage $V_{ac}$ across the electrodes, with d and S being the distance between the electrodes and the area of the electrode-plates. The values are, respectively, $V_{ac} = 0.5 - 10$ V, $\omega = 10^{-4} - 10^5$ Hz, $E_b = 0 - 5.0$ kV·mm$^{-1}$, $d = 0.1 - 0.2$ mm and $S = 1$ cm$^2$. The corresponding wave-vector $|k|$, therefore, is around $10^0$-$10^2$ cm$^{-1}$ if the fluid is treated as a homogeneous medium [1], but might be as large as $10^5$ cm$^{-1}$ when the nanostructures are considered.

*[Insert Figure 1 about here]*

Typical $\varepsilon'(\omega)$ results of an ER sample at different $E_b$'s are displayed in a $\varepsilon'$-$\omega$ plot (Fig. 1a). At $E_b = 0$, a large dispersion is observed to fall into two frequency regions (inset, Fig.1a). A Debye-like relaxation appears around $5\times10^3$ Hz with the $\varepsilon'$ decreasing from ~ 30 to ~ 10 between $10^2$ and $10^5$ Hz. It should be noted that the surface conductance of individual particles has been invoked to account for the similar features previously reported [8]. Below 1 Hz, however, both $\varepsilon'$ and $\varepsilon''$ increase as $\omega$ decreases, with $\varepsilon'$ rapidly reaching a value of 20,000 at ~ $10^{-4}$ Hz. They roughly obey the fractional power law of $\omega^{-n}$ with n $\approx$ 0.5, and the $\varepsilon'/\varepsilon''$-ratio is independent of $\omega$, as prescribed by the empirical 'universal dielectric response' (UDR) for a disordered solid, which can be modelled as a capacitor-resistor network [9,10]. Therefore, in the absence of an $E_b$, our samples behave similarly to other disordered systems at low $\omega$.

*[Insert Figure 2 about here]*

However, in the presence of an $E_b \geq 0.15$ kV·mm$^{-1}$, the situation changes drastically: $\varepsilon'$ decreases from its unbiased value and switches sign below ~ 1 Hz but with a negligible $E_b$-dependence at higher frequencies. As shown in Fig.1a, the positive $\varepsilon'$ drops precipitously to a negative value at a characteristic frequency $\omega_c$, which increases with $E_b$. For example, $\varepsilon'$ at $\omega \sim 10^{-3}$ Hz decreases from $+10^3$ to $-10^4$, when $E_b$ is increased from 0 to 0.5 kV·mm$^{-1}$. The $\omega_c$ increases from ~$3\times10^{-3}$ to $3\times10^{-1}$ Hz as $E_b$ rises from 0.5 to 3.5 kV·mm$^{-1}$. However, both $\varepsilon'(\omega)$ and $\omega_c$ change little for $E_b > 3.5$ kV·mm$^{-1}$. A complete set of the $\varepsilon'(\omega)$-data is shown in Fig. 2. A large low frequency dispersion in $\varepsilon''(\omega) = [\sigma(\omega) - \sigma(0)]/\omega$ is also observed. $\sigma(\omega)$ increases with $E_b$ and becomes ω-independent below a characteristic frequency $\omega_\sigma > \omega_c$, although both increase with $E_b$. When $\omega < \omega_\sigma$, $\sigma(\omega)$ is $E_b$-independent (Fig. 1b). A slight dip in $\sigma(\omega)$ is detected near $\omega_c$. While the $\sigma(\omega)$ can be understood in terms of the percolation paths associated with the field-induced closer particle-packing, a negative $\varepsilon'(\omega)$ is expected from neither the percolation nor the UDR model. Similar field effects on $\varepsilon'(\omega)$ and $\varepsilon''(\omega)$ are also observed in the pellet samples of U-BRTOCO nano-particles (Fig.1a), demonstrating that the silicone oil is not essential in the manifestation of negative $\varepsilon'$. Very recently, similar field-induced negative $\varepsilon'(\omega)$ has been detected [10] in another nano-particle colloid (but not with large particles) at $10^{-3}$ Hz, suggesting that field-induced $\varepsilon'$ sign-change is length-scale-sensitive and less material-sensitive.

While a field-induced negative $\varepsilon'$ at low $\omega$ has been reported previously in material systems such as Schottky diodes [11] and fuel cells [12], it had been considered to be a local electrochemical effect, thermal-induced de-trapping, or simply an experimental artefact. To determine $\varepsilon'$ at very low $\omega$, extra cautions have been taken. A technique-related artefact is not the cause since we employed ten different circuits and all yielded the same results. Calibrations against resistors and capacitors show that both the systematic phase-shifts as well as the experimental resolution of the phase-angle $\theta(\omega)=\tan^{-1}(\omega\varepsilon'/\sigma)$ are better than 0.05°, which is orders of magnitude smaller than our experimental values (inset, Fig.1b). The possibility of an electrode-related artefact is unlikely since the same $\varepsilon'(E_b)$ results are obtained when different electrodes of Cu, Ni, Pt, and Au are used. Opposite polarities of the applied $E_b$ produce the same negative $\varepsilon'$, effectively ruling out any local battery origin of the observation. In addition, the proposed electrochemical reactions and electrode effects should result in a large nonlinearity. Therefore, the non-linear effect was determined [10] by investigating the waveform of the out-of-phase current, the third harmonic, and the $V_{ac}$ and $E_b$-dependences of $\varepsilon'$. The non-linearity so-obtained below $\omega_c$ is not more than a few percent for $V_{ac}$ up to 100 V at $E_b > 2.5$ kV·mm$^{-1}$. No magnetic field greater than $10^{-6}$ T has been detected in our samples with a negative $\varepsilon' \sim -10^5$, ruling out any inductive origin of the phenomenon. It is known that particles of similar charge attract in a material with a negative ε'(ω), which will lead to instabilities, such as the formation of charge density waves. However, the observed noise spectrum follows the standard 1/ω–dependence, and displays no anomaly at $\omega_c$, indicating that no electronic instability is involved. The arrest of such instabilities may be the result of the $E_b$ applied. All of the above demonstrate that the field-induced negative ε' is indeed an intrinsic property of the nano-particle aggregates.

In a system exhibiting a resonance, a negative ε'(ω) occurs at frequencies slightly higher than the resonant frequency $\omega_o$. The nano-particles in our ER fluid sample are known [6] to form columns in the presence of $E_b$. A possible electric-field-driven mechanical resonance at $\omega_o$ in such column-structures is examined. However, the estimated $\omega_o = \sqrt{(k/m)} \sim 10^5$ Hz, where $k$ and $m$ are respectively the elastic modulus and the mass of the assembly, is much greater than the $\omega_c$ $\sim 10^{-1}$ Hz observed. This is further confirmed by the observation of a negative ε'(ω) with a low-$\omega_c$ in the pellet sample, where mechanical oscillation, if it exists, is expected to occur at an even higher frequency. On the other hand, the free plasma model [13], $\varepsilon'(\omega) = 1 - \omega_p^2/(\gamma^2 + \omega^2)$, accounts for the data well (Fig.1a), where $\omega_p$ is the plasma frequency and γ is the damping parameter. The γ deduced is rather small with a value $\sim 10^{-4}$ s$^{-1}$. The small γ implies that the carriers will keep their initial phase intact for hours without significant interference from particle collisions, thermal excitation, and electromagnetic noise. This suggests that new collective excitations may be the origin of the negative ε'(ω). Unfortunately, the $\omega_p = \sqrt{(Ne^2/m)} \sim 10^{-1}$ Hz in the free plasma model so obtained, where $N$, $e$, and $m$ are the density, charge, and mass of the carriers, respectively, leads to a very low carrier density of $10^4 - 10^6$/cm$^3$, whose physical meaning remains a puzzle. However, the impasse can be alleviated to a large extent by increasing the effective mass of the carriers. The observed $\sigma(\omega) = \sigma(0) + \omega\varepsilon''(\omega)$ includes the *dc* percolative conductance and thus prevents a direct comparison with the model.

*[Insert Figure 3 about here]*

We want to determine if a negative polarization *P* exists in the U-BRTOCO nano-particle aggregates in the presence of a positive field. The *I*(*V*)-loops of the sample have therefore been measured over a *V*-range of ±200 V (or field-range of ±2 kV·mm$^{-1}$) at frequencies of 1 – 10$^{-4}$ Hz. At 1 Hz, a singly connected clockwise loop centred on the origin is detected, as expected of an ordinary capacitor. However, at lower ω, two counter-clockwise loops appear at high |*V*| (inset, Fig. 3), in addition to the clockwise one around the origin (too small to be resolved with the scale used in the inset of Fig. 3), suggesting the reversal of the I-sign at high |*V*|. To investigate the charge *Q* accumulated across the sample, we have integrated the hysteretic component of *I*, i.e. $Q(t) = \int_0^t (I - I_{reversible}) d\tau$, which is proportional to *P* and shown together with the *V* as a function of time. *Q* or *P* is found to be always negative under positive *V*, except for a narrow *V*-range within ± 15 V. The negative ε' and the opposite directions of the local field and the induced polarization in the U-BRCOTO nano-particle aggregates appear to be the expected properties of the missing dielectric analogue of a diamagnet, or 'diaelectricity'. However, it should be noted that the 'diaelectricity' observed here is accompanied with considerable resistive loss. Other possible 'diaelectric' properties are being sought.

**Acknowledgements**

The work in Houston is supported in part by US National Science Foundation Grant No. DMR-9804325, the T. L. L. Temple Foundation, the John J. and Rebecca Moores Endowment, the Strategic Partnership for Research in Nanotechnology and the State of Texas through the Texas Center for Superconductivity at the University of Houston; and at Lawrence Berkeley Laboratory by the Director, Office of Science, Office of Basic Energy Sciences, Division of Materials Sciences and Engineering of the U.S. Department of Energy under Contract No. DE-AC03-76SF00098.

* Correspondence and requests for materials should be addressed to C.W.C. (e-mail: cwchu@uh.edu).

**FIGURE CAPTIONS**

Figure 1: The field-induced dielectric constant sign-switch in nano-particle aggregates: a) Typical electric field effect on $\varepsilon'(\omega)$ for different $E_b$ in kV·mm$^{-1}$: 0 - ●, 0.5 - ▼, 2.0 - ◆, 3.5 - ▲ and 5.0 - ✕ for the ER-sample, and 0 - ○ and 0.36 - ★ for the pellet sample. Inset – $\varepsilon'(\omega)$ at $E_b = 0$. b) Electric field effect on $\sigma(\omega)$ for the ER-sample using the same symbols as in Fig. 1a. Inset – the phase-angle θ at different ω and $E_b$ with the experimental resolution represented by the width of the centred horizontal line.

Figure 2: The $\varepsilon'(\omega)$-log(ω)-$E_b$ three-dimensional plot for the nano-particle aggregates.

Figure 3: The opposite signs of the field and the accumulated charge in the nano-particle aggregates at high fields, but same sign for small fields as bracketed by the vertical dashed lines: $V(t)$ and $Q(t)$. Inset – $I(V)$ loops.

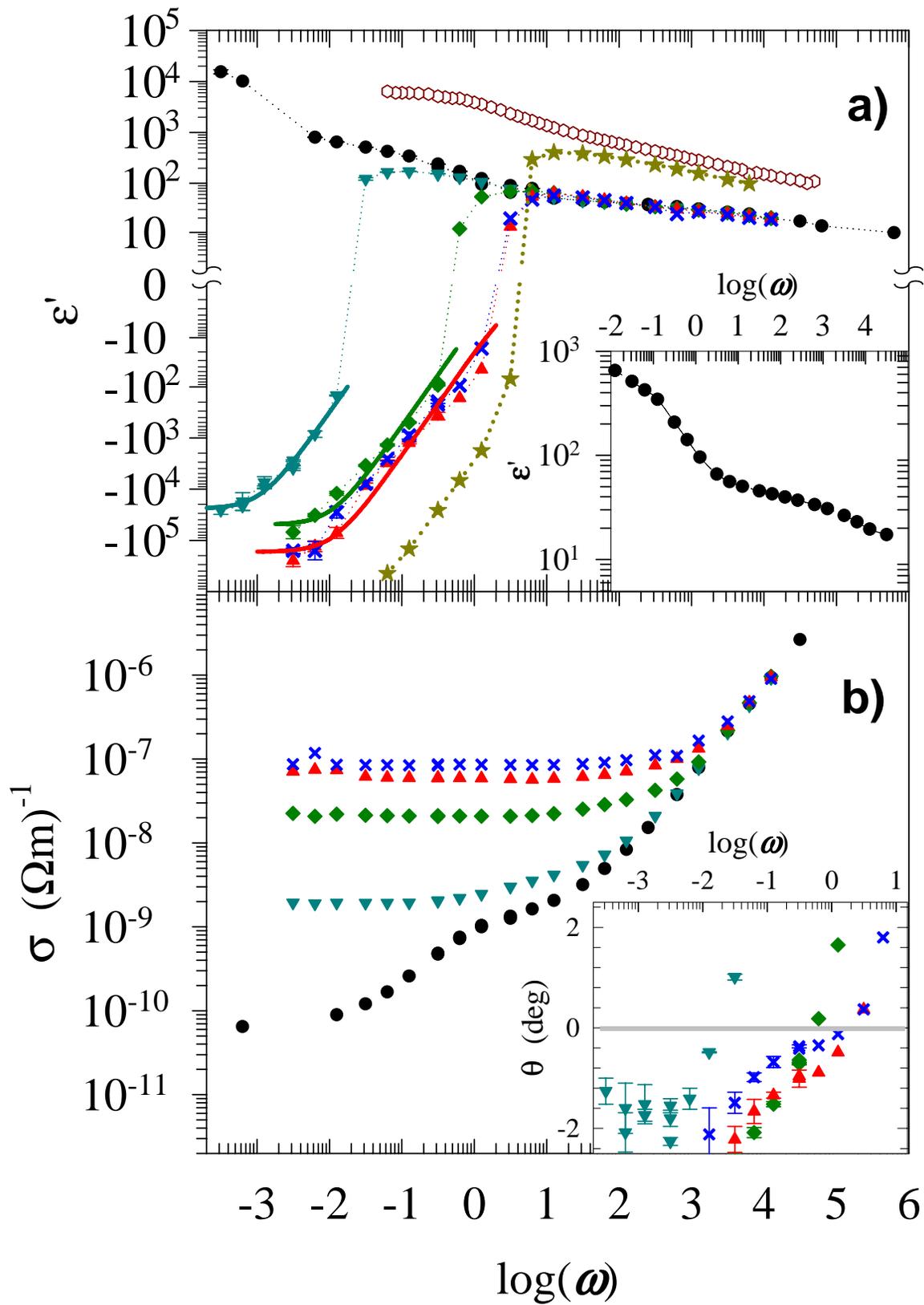

Figure 1, Chen *et al*

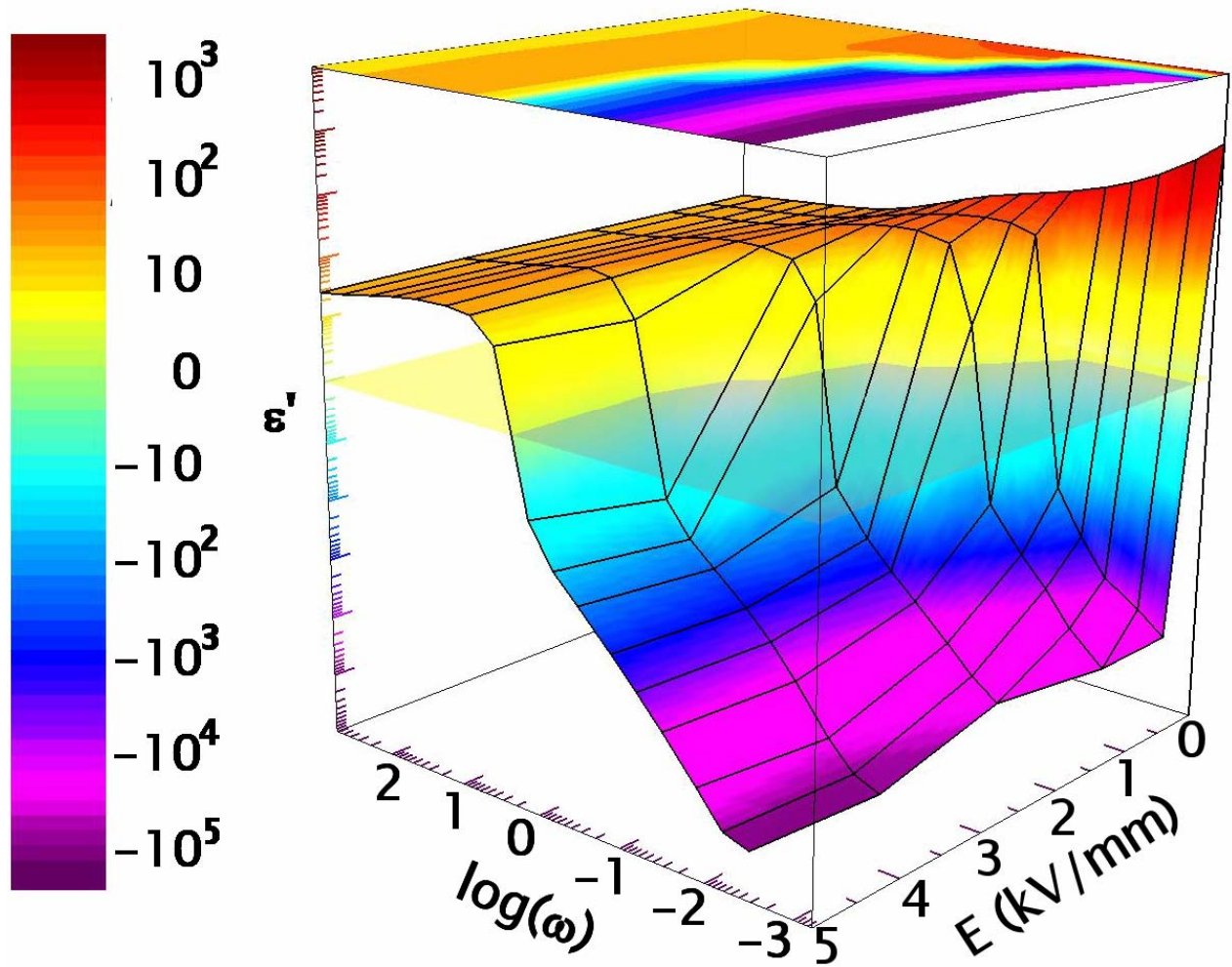

Figure 2, Chen *et al*

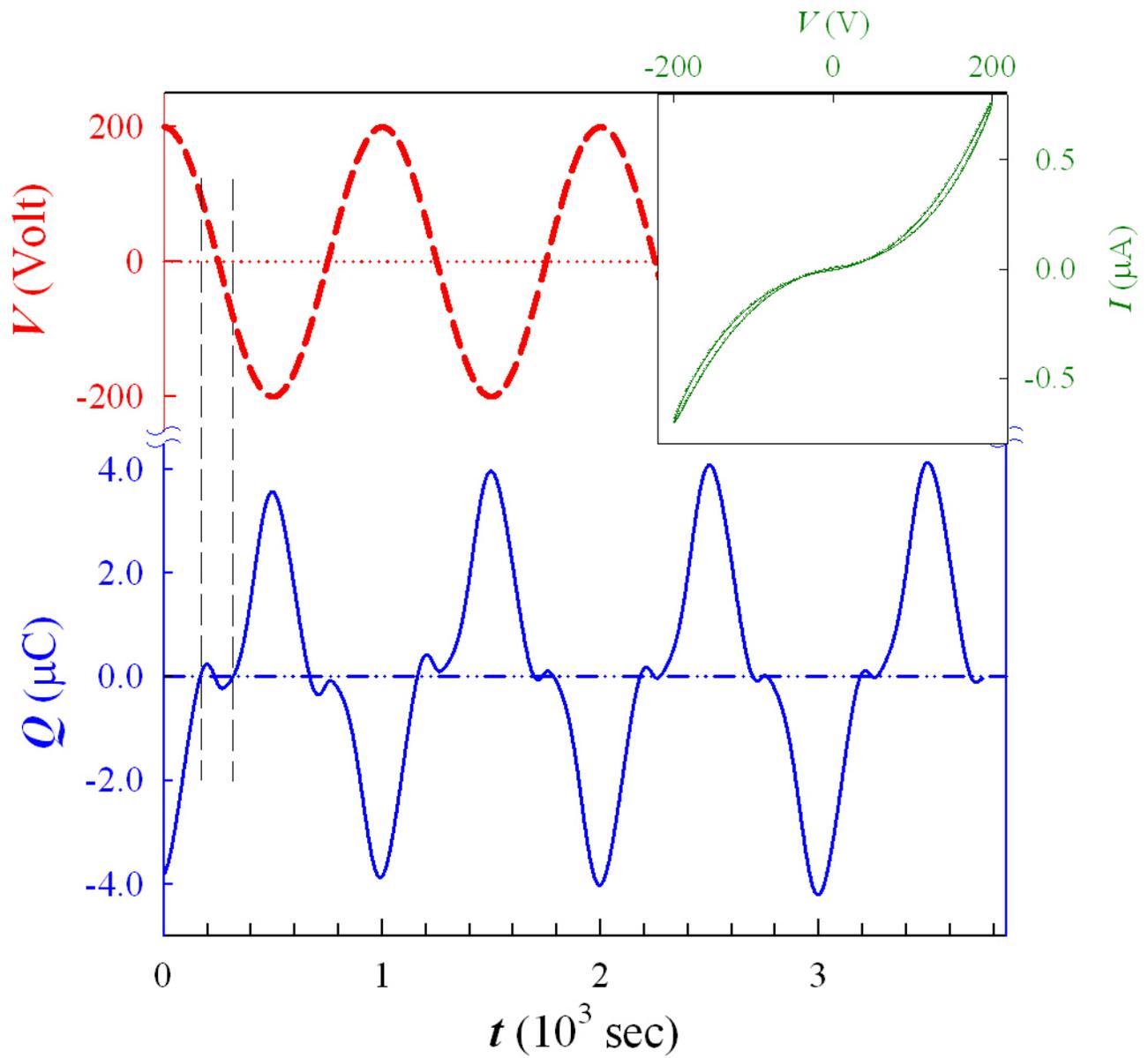

Figure 3, Chen *et al*